\documentclass[letterpaper,usenatbib]{mnras}
\usepackage{graphicx}
\usepackage{multirow}
\usepackage{color}
\usepackage{lscape}
\usepackage[figuresright]{rotating}
\usepackage{longtable}
\usepackage{mathrsfs,amsmath}
\usepackage{epstopdf}
\newsavebox{\tablebox}
\usepackage{hyperref}
\usepackage{times}
\usepackage{longtable,lscape}

\newcommand{\lv}{\ifmmode L_{5100} \else $L_{5100}$\ \fi}
\newcommand{\kms}{\ifmmode {\rm km\ s}^{-1} \else km s$^{-1}$\ \fi}
\newcommand{\ergs}{\ifmmode {\rm erg\ s}^{-1} \else erg s$^{-1}$\ \fi}
\newcommand{\lb}{\ifmmode L_{\rm Bol} \else $L_{\rm Bol}$\ \fi}
\newcommand{\ledd}{\ifmmode L_{\rm Edd} \else $L_{\rm Edd}$\ \fi}
\newcommand{\hb}{\ifmmode H\beta \else H$\beta$\ \fi}
\newcommand{\ha}{\ifmmode H\alpha \else H$\alpha$\ \fi}
\newcommand{\mbh}{\ifmmode M_{\rm BH}  \else $M_{\rm BH}$\ \fi}
\newcommand{\msun}{M_{\odot}}
\newcommand{\rfe}{\ifmmode R_{\rm Fe} \else $R_{\rm Fe}$\ \fi}
\newcommand{\sst}{\ifmmode \sigma_{\rm \ast}\else $\sigma_{\rm \ast}$\ \fi}
\newcommand{\dhb}{\ifmmode D_{\rm H\beta} \else $D_{\rm H\beta}$\ \fi}
\newcommand{\leddR}{\ifmmode L_{\rm Bol}/L_{\rm Edd} \else $L_{\rm Bol}/L_{\rm Edd}$\ \fi}
\newcommand{\feii}{Fe {\sc ii}\ }
\newcommand{\mdot}{\ifmmode \dot{\mathscr{M}}  \else $\dot{\mathscr{M}}$\ \fi}
\newcommand{\rhb}{\ifmmode R_{\rm BLR}({\rm H\beta})  \else $R_{\rm BLR}({\rm H\beta})$ \ \fi}
\newcommand{\shb}{\ifmmode \sigma_{\rm H\beta} \else $\sigma_{\rm \hb}$\ \fi}
\newcommand{\RL}{\ifmmode R_{\rm BLR}({\rm H\beta}) - L_{\rm 5100} \else $R_{\rm BLR}({\rm H\beta}) - L_{\rm 5100}$ \ \fi}
\newcommand{\ms}{\ifmmode M_{\rm BH}-\sigma_{\ast} \else $M_{\rm BH}-\sigma_{\ast}$\ \fi}
\begin{document}
\title[An extended size-luminosity relation]{An extended size-luminosity relation for the reverberation-mapped AGNs: the role of the accretion rate}
\author[L. Yu et al.]{Li-Ming Yu$^{1}$, Bi-Xuan Zhao$^{1}$, Wei-Hao Bian$^{1}$ \thanks{E-mail: whbian@njnu.edu.cn}, Chan Wang$^{1}$, Xue Ge$^{1}$ \\
$^1$School of Physics and Technology, Nanjing
Normal University, Nanjing 210046, China\\
}\maketitle

\begin{abstract}
For a compiled sample of  120 reverberation-mapped AGNs, the bivariate correlations of the broad-line regions (BLRs) size ($R_{\rm BLR}$) with the continuum luminosity at 5100 \AA\ (\lv) and the dimensionless accretion rates (\mdot) are investigated. Using our recently calibrated virial factor $f$, and the velocity tracer from the \hb Full-width at half-maximum (FWHM(\hb)) or the line dispersion ($\sigma_{\rm \hb}$) measured in the mean spectra, three kinds of SMBH masses and \mdot are calculated. An extended \RL relation including \mdot is found to be stronger than the canonical \RL relation, showing smaller scatters. The observational parameters, \rfe (the ratio of optical \feii to \hb line flux) and the line profile parameter \dhb ($\dhb=\rm FWHM(\hb)/\sigma_{\rm \hb}$), have relations with three kinds of \mdot. Using \rfe and \dhb to substitute \mdot, extended empirical \RL relations are presented. 
\rfe is a better "fix" for the \RL offset than the \hb shape \dhb. The extended empirical \RL relation including \rfe can be used to calculate $R_{\rm BLR}$, and thus the single-epoch SMBH mass \mbh.
Our measured accretion rate dependence is not consistent with the simple model of the accretion disk instability leading the BLRs formation. The BLR may instead form from the inner edge of the torus, or from some other means in which BLR size is positively correlated with accretion rate and the SMBH mass.
 %The accretion rate has an important role in the \RL relation. 
\end{abstract}


\section{INTRODUCTION}
There is good observational and theoretical evidence that supermassive black holes (SMBHs) exist in nearly every galaxy in universe. Understanding the properties of these SMBHs will clarify their roles in galaxy formation and evolution across the cosmology history \citep[e.g.][]{KH13}. There are mainly two parameters for a SMBH, i.e., mass (\mbh) and spin, which need to be determined. For a few very nearby ($<$ 100 Mpc) quiescent galaxies, including our Galaxy, SMBH masses can be measured through the stellar or gaseous dynamics method \citep[e.g.][]{Tr02,Mc11}. It has been found that nearby quiescent galaxies follow a tight correlation between the central SMBH mass and the bulge or spheroid stellar velocity dispersion (\sst), which is called \ms relation \citep[e.g.][]{KH13}. AGNs can be classified into type 1 or type 2 AGNs, depending on whether the broad-line regions (BLRs) can be viewed directly.  For type 1 AGN, the BLR can be used as a probe of the gravitational potential of the SMBHs. The SMBH mass can be weighed through the BLRs clouds for type I AGNs across cosmos time. The SMBH masses in type I AGNs can be calculated as follows \citep[e.g.][]{Ka00, bian2002, Pe04, Co06, Du16a, Yu19}:

 \begin{equation}
 \label{eq1}
\mbh=f_{\rm BLR}\frac{R_{\rm BLR}~(\Delta V)^2}{G}.
\end{equation}
where  $G$ is the gravitational constant. $R_{\rm BLR}$ is the distance from black hole to the BLRs,  and can be estimated from the reverberation mapping (RM) method \citep[e.g.][]{BM82,Pe93}.  $\Delta V$ is the velocity of the BLRs clouds, and usually traced by the Full-width at half-maximum (FWHM) or the line dispersion ($\sigma_{\rm \hb}$) of the broad \hb emission line. $f_{\rm BLR}$ is a virial factor to characterize the kinematics, geometry, inclination of the BLRs clouds. Using the \ms relation, we recently did the calibration of $f_{\rm BLR}$ and found $f_{\rm BLR} \propto \rm FWHM^{-1.11}$ when FWHM(H$\beta$) is used as the tracer of $\Delta V$ in Equation \ref{eq1} \citep{Mejia2018,Yu19}. It is consistent with the results by the BLRs dynamical model to fit simultaneously the AGNs continuum/\hb light curves and \hb line profiles \citep[e.g.][]{Li2018, Pancoast2018, Williams2018}.

To weigh SMBH masses in AGNs, $R_{\rm BLR}$ is a key parameter in Equation \ref{eq1}. Considering the photonionization model of the BLRs in AGNs, the variance of the central ionization luminosity leads to the variability of the emission line luminosity in the BLRs, but with a time lag $\tau$ (i.e. the RM technique). $R_{\rm BLR}=c\tau$ can be estimated, and the lag time $\tau$ was successfully measured by the RM method for nearly 120 AGNs \citep[e.g.][]{Pe04, Gr17, Fau2017, Du18}. An empirical \RL relation for the \hb broad line is derived based on the RM AGNs and is also suggested in a single AGN. \citep{Ka00, Ka05, Be13, KE15, Du18}:
 \begin{equation}
 \label{eq2}
R_{\rm BLR}({\rm \hb}) = \alpha~ l_{44}^{\beta}~~\, \rm ltd.
\end{equation}
where $l_{44} = \lv/10^{44}\ergs$ is the 5100 \AA\ luminosity in units of $10^{44}$ \ergs and $R_{\rm BLR}(\rm \hb) = c\tau$ is the emissivity-weighted radius of the BLRs \citep{Ka00, Be13}. This empirical \RL relation is widely calibrated and used to calculate the single-epoch \mbh for AGNs showing broad emission-lines from a flux-calibrated spectrum \citep[e.g.][]{bian2004, VP06, bian2008, Sh11, Ge16}. 
%Assuming $\beta=0.5$, the formula for the single-epoch \mbh was given by \cite{VP06}. 
Subtracting the starlight contribution at 5100 \AA\ by decomposing the two-dimensional surface brightness of the host galaxies of RM AGNs imaged by HST, \cite{Be13} found that 
$\beta=0.533\pm 0.03$ instead of previous value of $0.70\pm 0.03$ by \cite{Ka00}. 

For a large RM campaign of super-Eddington accreting massive black holes (SEAMBHs) in AGNs \citep{Wang2013}, it was found that the \hb time lags of the SEAMBHs are shorter than the values predicted by the canonical \RL relation of sub-Eddington AGNs, by factors of $\sim 2-6$, and the \hb size for super-Eddingon AGNs (\mdot $\ge$ 3) has a dependence on the dimensionless accretion rate \mdot \citep[e.g.][]{Du18, Ma2019}. 
For a high-z sample of 44 RM AGNs ($z  ~\sim 0.1-1.0$) from the Sloan Digital Sky Survey (SDSS) RM Project, \cite{Gr17} measured \hb/\ha lags and also found shorter lags for a number of AGNs. The shorter lag is possibly due to the self-shadowing effect in super-Eddington AGNs, the retrograde accretion onto the SMBH in low-accretion rate AGNs\citep{Wang2014a, Wang2014b, Du18}. 
Recently, it was suggested that the relation between \lv and the ionizing flux $L_{\rm ion}$ is non-linear and depends on the Spectral Energy Distribution (SED) of the source, which would lead to a shorter time lag \citep{C2019}. 
Considering that the gravitational instability of standard thin accretion disks leads to the BLRs, it was suggested that the BLRs sizes also have a relation with the mass accretion rate ($\dot{M}$) in addition to the continuum luminosity at 5100 \AA, $R_{\rm  BLR} \propto L_{\rm 5100}^{0.5} \dot{M}^{-37/45}$ \citep{bian2002}. Therefore, from both sides of RM observation or the BLRs theory, it needs to revise the empirical \RL relation to consider the effect of the SMBH accretion rate.

The accretion rate can be derived  from the disk model of \cite{SS73}, which has been extensively applied to fit the spectra of AGNs\citep[e.g.][]{DL2011, Wang2014a}. Considering that the effective temperature distribution is given by $T_{\rm eff} \propto R^{-3/4}$ and the effect of the inner boundary can be neglected because the region emitting optical radiation is far from the boundary, the dimensionless accretion rate \mdot is
 \begin{equation}
 \label{eq3}
\mdot \equiv \dot{M}/\dot{M}_{Edd} = 20.1\left(\frac{l_{44}}{\cos\,\it{i}}\right)^{3/2} m_{7}^{-2}.
\end{equation}
where $l_{44} = \lv/10^{44}\ergs$ is the 5100 \AA\ luminosity in units of $10^{44}$ \ergs, the Eddington accretion rate is $\dot{M}_{\rm Edd}\equiv \ledd / c^2$, the Eddington luminosity is $\ledd =1.5\times 10^{38} (\mbh/\msun)$, $i$ is the accretion disk inclination relative to the observer, the SMBH mass $m_{7}=M_{\rm BH}/10^7 \msun$. An average value of $\rm cos~ i = 0.75$ is adopted 
\citep[e.g][]{Du18}.
%, which corresponds to the opening angle of the dusty torus. 
The Eddington ratio \lb/\ledd is an important parameter describing the SMBH accretion process, where \lb is the bolometric luminosity. $\lb=k_{5100} \times \lv$, and $k_{5100}=9$. The bolometric correction coefficient $k_{5100}$ was also suggested to have a dependence on the luminosity or the Eddington ratio \citep[e.g.][]{Marconi2004, Jin2012, Wang2019}. For the mass accretion rate $\dot{M}$, $\lb=\eta \dot{M} c^2$, then $\lb/\ledd=\eta \mdot$, where $\eta$ is the radiative efficiency. If $\eta=0.1$, then $\mdot=10\times \lb/\ledd$. 
For SEAMBHs, smaller $\eta$ leads to larger \mdot \citep[e.g.][]{Du18}.

It was suggested that dimensionless observational parameter \rfe or \dhb from the optical spectrum of AGNs can be used to determine the Eddington ratio \mdot, where the shape of the \hb broad-line profile $\dhb=\rm FWHM(\hb)/\sigma_{\hb}$, \rfe is the ratio of optical \feii to \hb line flux \citep[e.g.][]{Du16a}. The value of \dhb is 2.35, 3.46, 2.45, 2.83, and 0 for a Gaussian, a rectangular, a triangular, an edge-on rotating ring, and a Lorentzian profile, respectively. \dhb depends on the line profile and gives a simple, convenient parameter that may be related to the dynamics of the BLRs \citep{Co06, Du16a}. \rfe has a correlation with Eigenvector 1 (EV1) from principal component analysis, which has been demonstrated to be driven by the Eddington ratio \citep{BG92, Su00, Boroson2002, Sh14, Ge16}.

In this paper, for a large compiled sample of 120 RM AGNs, considering the role of the dimensionless accretion rate \mdot, the bivariate correlation of $R_{\rm BLR}$ with \lv and \mdot is investigated. 
%An extended empirical \RL relation including observational parameter of \rfe or \dhb is presented. 
Section 2 presents our sample. Section 3 is data analysis and Section 4 is discussion. Section 5 summaries our results. 
All of the cosmological calculations in this paper assume $\Omega_{\Lambda}=0.68$, $\Omega_{\rm M}=0.32$, and $H_{0}=67~ \kms {\rm Mpc}^{-1}$ \citep{Plank2014}.

\section{SAMPLE}
Up to now, there are 120 RM AGNs with measured \hb lags \citep[e.g.][]{Be13,Du15,Gr17,Du18}. We divide them into three subsamples. The first subsample has 25 AGNs presented by SEAMBH collaboration (hereafter SEAMBHs) \citep{Du15,Du16b,Du18}. These 25 sources were observed by SEAMBH collaboration since 2012. 24 out of 25 are identified to be super-Eddington accretor ($\mdot \ge 3$) \citep{Du18}. MCG+06-26-012 was selected as a super-Eddington candidate in SEAMBH2012 but was later identified to be a sub-Eddington AGN (\mdot = 0.46). The second subsample contains 39 AGNs summarized by \cite{Be13} (excluding Mrk 335 and Mrk 142 again mapped by the SEAMBHs project) and 12 other sources published recently (hereafter BentzSample) \citep{Ba15, Be16a, Be16b, Fau2017, Williams2018}. The third subsample contains 44 high-z AGNs ($z \sim 0.1 - 1.0$) from SDSS-RM project which was done by fibre spectrum (herafter SDSS-RM) \citep{Gr17}. The entire sample includes 120 AGNs with measured \hb lags. We collect \dhb and \rfe measured from the optical spectrum, which were suggested to be driven by \mdot.

Properties about our sample of 120 AGNs are shown in Table \ref{tab1}. Col. (1) is the source name. Col. (2) is the rest-frame \hb lag in units of days. Col. (3) is host-corrected monochromatic luminosity at 5100 \AA. Col. (4) is the broad \hb FWHM measured from the mean spectrum. Col. (5) is the broad \hb line dispersion \shb measured from the mean spectrum. Col. (6) is \dhb and Col.(7) is \rfe.  Col(14) is the reference for these data \citep[e.g.][]{Ba13, Ba15, Be06, Be09a, Be09b, Be13, Be14, Co06, Denny2006, Denny2010, Du15, Du16a,Du16b, Du18, Fau2017, Gr12, Gr17, HK14, Lu16, Pei14, Pei17, Pe00, Pe14, Sh15, Shen19, Williams2018, Zhang19}. The \lv of SDSS-RM sources comes from \cite{Sh15} and the \rfe is derived from \cite{Shen19}, Col. (2), Col. (4) and Col. (5) come from \cite{Gr17}. For the other sources, Col.(2)-Col. (5), Col.(7) and Col.(11) are mainly from \cite{Du15} and \cite{Du16a}.  For a source with multiple RM measurements, we use the square of measurement error as weight to calculate the weighted average (called the "averaged scheme"). The error of the compiled data is calculated from the weighted average of the measurement errors and the weighted standard deviation. We use the averaged scheme in our analysis like that in \cite{Du15,Du18}. 

\section{Data Analysis}

\subsection{The SMBH mass \mbh and the dimensionless accretion rate \mdot}

\begin{figure*}
\includegraphics[angle=-90,width=6.0in]{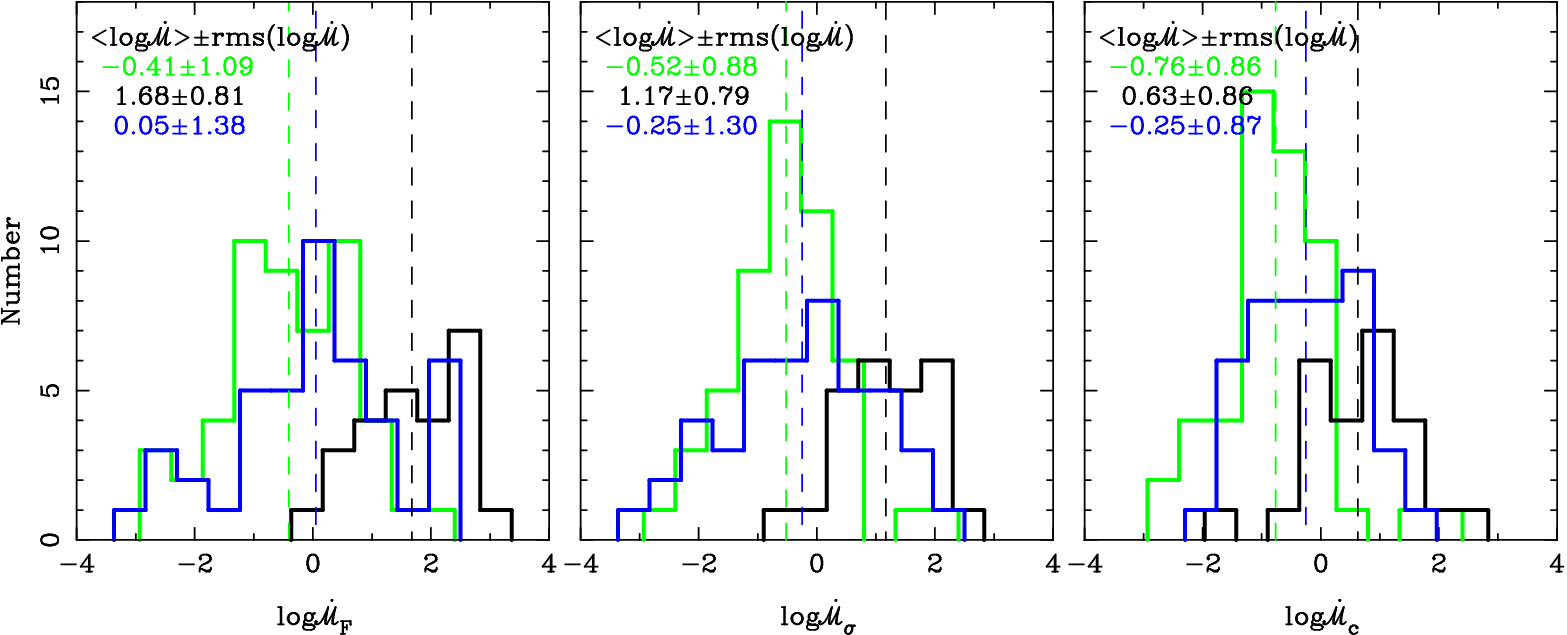}
\caption{Distributions of three kinds of \mdot, i.e., $\mdot_{\rm F}, \mdot_{\rm \sigma}, \mdot_{\rm c}$ from left to right, respectively. The black, green, blue lines denote the subsamples of SEAMBH, BentzSample, SDSS-RM,  respectively. The dash line is the mean value for these three kinds of \mdot. The mean value and the rms are shown in the upper left corner of each panel.}
\label{fig1}
\end{figure*}

\begin{figure*}
\includegraphics[angle=-90,width=6.0in]{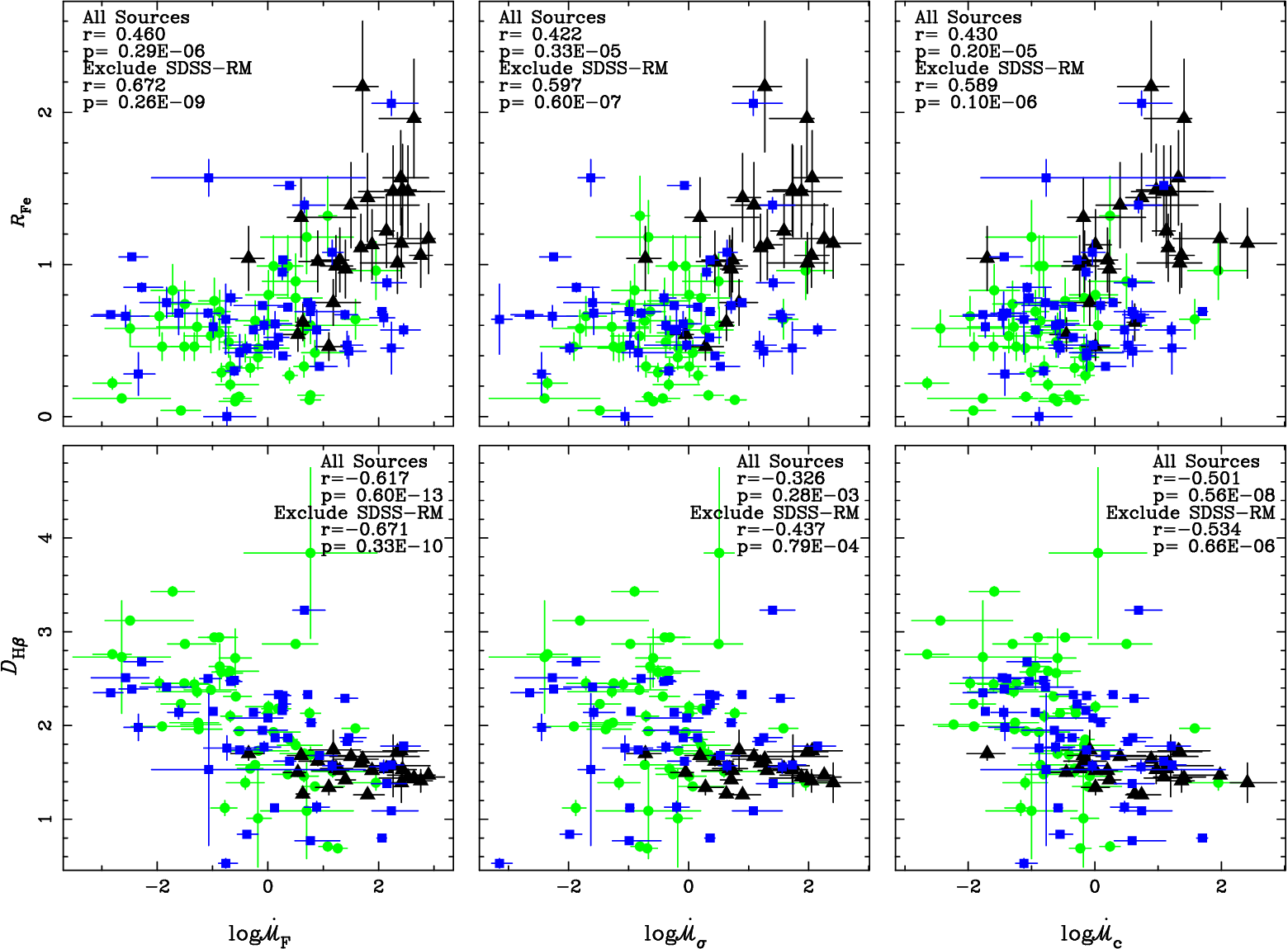}
\caption{Correlations of \mdot with \rfe (top) and \dhb (bottom). In the left two panels, the dimensionless accretion rate $\mdot_F$ is derived from FWHM and the virial factor $f_{\rm BLR}$=1. In the middle two panels, $\mdot_{\sigma}$ is derived from \shb and $f_{\rm BLR}$=4.47. In the right two panels, $\mdot_c$ is derived from FWHM and the FWHM-based factor $f_{\rm BLR} \propto \rm FWHM^{-1.11}$. The source in the subsample of SEAMBHs is shown as a black triangle. The source in the subsample of BentzSample is shown as a green circle. The source in the subsample of SDSS-RM is shown as a blue square. The correlation coefficient $r_s$ and the null possibility $p_{\rm null}$ are shown in the left/right corner in each panel. }
\label{fig2}
\end{figure*}

\begin{figure*}
\includegraphics[angle=-90,width=6.0in]{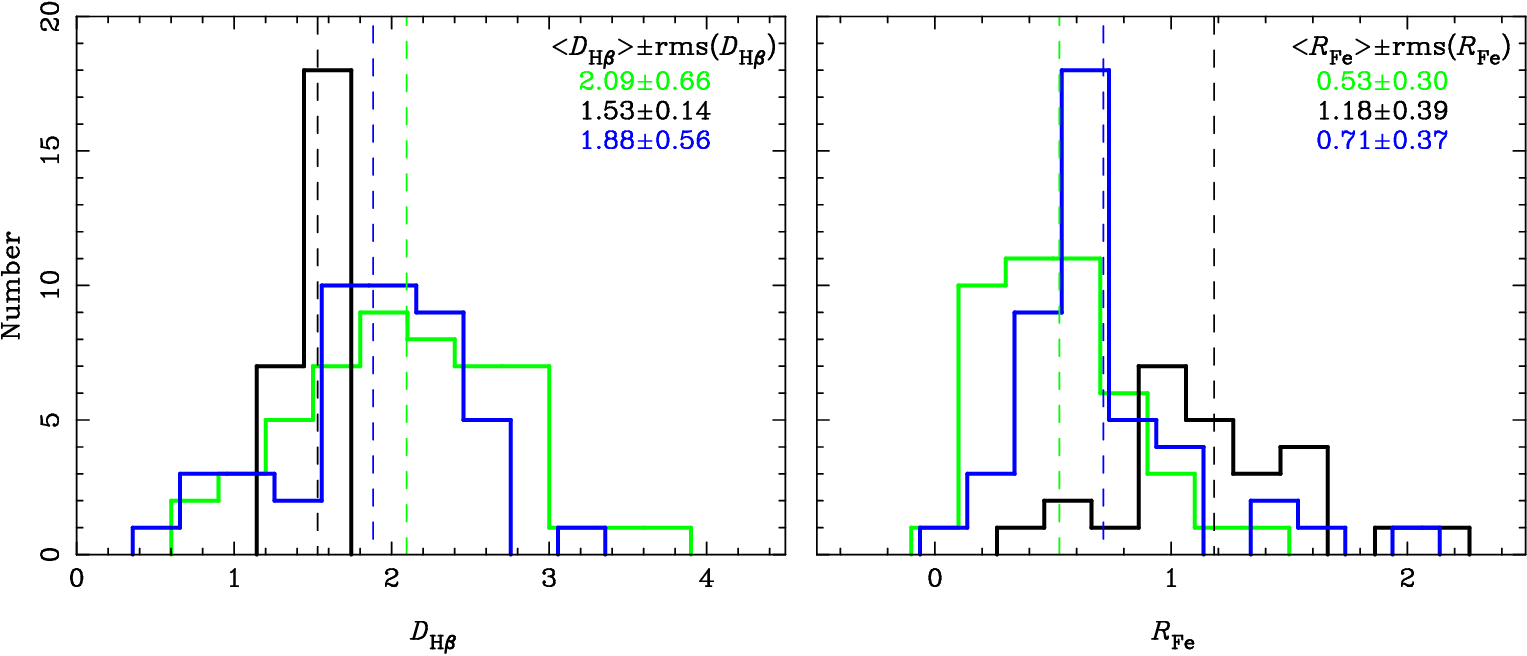}
\caption{The distribution of \dhb (left) and \rfe (right) for three subsamples. The mean value and the rms are shown in the upper left corner of each panel. The  black, green, blue lines denote the subsamples of SEAMBHs, BentzSample, and SDSS-RM, respectively.}
\label{fig3} 
\end{figure*}

We use this sample of 120 RM AGNs with measured \hb lags to do the bivariate correlation analysis to investigate the effect of \mdot in the \RL relation. About the downward offset for SEAMBH and SDSS-RM AGNs with respect to the canonical \RL relation, the dimensionless accretion rate \mdot is suggested to be the key parameter \citep{Du16b, Du18}. Therefore, we use \mdot and \lv to do the bivariate fitting. In this paper, we don't use the Eddington ratio \lb/\ledd. Given the uncertainty in bolometric corrections, we do not attempt to infer \lb from the monochromatic luminosity at 5100 \AA\
\citep[e.g.][]{Marconi2004, Trump2011, Jin2012, Wang2019}.

The SMBH masses \mbh of 120 broad-line RM AGNs can be derived from the Eq. \ref{eq1}. Here we use three methods to derive \mbh and therefore yield three kinds of \mdot. Using the FWHM(\hb) from the mean spectrum to trace the velocity of the BLRs clouds and adopting the virial factor $f_{\rm BLR}$ as 1 \citep[e.g.][]{Du16b,Du18}, we denote this SMBH mass as $M_{\rm BH, F}$. Using the \shb instead of FWHM(\hb) from the mean spectrum to trace the velocity of the BLRs clouds and adopting $f_{\rm BLR}$ as 4.47 \citep{Wo15}, we denote this SMBH mass as $M_{\rm BH, \sigma}$. It was found that the line dispersion $\sigma_{\rm \hb}$ is insensitive to the inclination, while \hb FWHM has some dependence on the inclination \citep{Co06,Yu19}. In our previous work \citep{Yu19}, we proposed a new calibration about the virial factor, the FWHM(\hb)-based virial factor, has an anti-correlation with the FWHM of the \hb broad component:
\begin{equation}
\label{eq4}
\log f_{\rm BLR}= -1.11\log\frac{\rm FWHM(\hb)}{2000~\kms}+0.48.
\end{equation}
We denote this SMBH mass as $M_{\rm BH, c}$. Therefore we can use three kinds of the \mbh to derive \mdot, we denote these corresponding \mdot as $\mdot_{\rm F}$, $\mdot_{\rm \sigma}$ and $\mdot_{\rm c}$. $\mdot_{\rm F}$ is adopted as the dimensionless accretion rate in the SEAMBHs project \cite[e.g.][]{Du16a,Du16b, Du18}. In Table \ref{tab1},  Col. (8) is the black hole mass $M_{\rm BH, F}$ derived from the FWHM and $f_{\rm BLR}=1$.  Col. (9) is $M_{\rm BH, \sigma}$ derived from \shb and  $f_{\rm BLR}=4.47$ suggested by \cite{Wo15}. Col. (10) is $M_{\rm BH, c}$ derived from FWHM and $f_{\rm BLR} \propto \rm FWHM^{-1.11} $ \citep{Yu19}. For some AGNs, we use the MCMC (Markov chain Monte Carlo) \mbh from the BLR dynamical method, where the virial factor $f_{\rm BLR}$ is directly constrained from velocity-resolved reverberation mapping. \citep[e.g.][]{Li2018, Pancoast2018, Williams2018} (see table 6 in \cite{Yu19}). Columns (11-13) are the corresponding dimensionless accretion rates \mdot. From Equations \ref{eq4} and \ref{eq3}, assuming $\rm FWHM(\rm \hb)=5413~ \kms$, then $f_{\rm BLR}=1$, and $M_{\rm BH, c}$ is the same as $M_{\rm BH, F}$ for a source. Assuming $\rm FWHM(\rm \hb)=2000~ \kms$, then $f_{\rm BLR}=3.02$ and $M_{\rm BH, c}$ is larger than $M_{\rm BH, F}$ by 0.48 dex, $\mdot_c$ is smaller than $\mdot_F$ by 0.96 dex for a source. 

%Therefore, the classification of super-Eddington AGNs depends on the calculation of \mbh.   

In Fig. \ref{fig1}, we show the distributions of three kinds of \mdot. The black line is for the first subsample of SEAMBHs AGNs. The green line is for the second subsample of BentzSample. The mean value and the rms of $\log\mdot$ are shown in each panel. The blue line is for the third subsample of SDSS-RM AGNs. The subsample of SEAMBHs has the largest mean value of \mdot which is expected due to the selection criteria of the sample. The subsample of SDSS-RM has the larger distribution range than the others, which is similarly expected from SDSS sample selection \citep{Sh15}. For $\log\mdot_{\rm F}$, the mean values with the standard deviations in distributions are $1.68\pm 0.81$, $-0.41\pm 1.09$, $0.05\pm 1.38$,  respectively for the subsamples of SEAMBHs, BentzSample, SDSS-RM. For $\log\mdot_{\sigma}$, they are respectively $1.17\pm 0.79$, $-0.52\pm 0.88$, $-0.25\pm 1.30$. For $\log\mdot_{\rm c}$, they are respectively $0.63\pm 0.86$, $-0.76\pm 0.86$, $-0.25\pm 0.87$.
Adopting our FWHM-based $f$ formula (Equation \ref{eq4}), for subsample of SEAMBHs, smaller FWHM leads to larger $f_{\rm BLR}$. It would lead to larger \mbh and thus smaller \mdot than for adopting $f_{\rm BLR}=1$. Adopting the super-Eddington criterion of $\mdot \ge 3$ \citep{Du18}, using $\mdot_{\rm F}$, there are 24/25, 14/51, 16/44 super-Eddington SMBHs in three subsamples, respectively. Using $\mdot_{\rm \sigma}$, there are 20/25, 6/51, 13/44 super-Eddington SMBHs, respectively. Using $\mdot_{\rm c}$, there are 14/25, 3/51, 12/44 super-Eddington SMBHs, respectively. With $\mdot_{\rm c}$, the number of super-Eddington sources become smaller. Adopting FWHM-based $f$ in $\mdot_{\rm c}$, smaller FWHM(\hb) leads to larger $f$, and then larger \mbh, which makes some AGNs being sub-Eddington. The classification of super-Eddington AGNs depends on the calculation method of \mbh and then \mdot (see Table 1 \ref{tab1}).

%\subsection{The correlations of \mdot with \dhb and \rfe}

In Fig. \ref{fig2}, we find that \mdot has correlations with \dhb and \rfe, which is consistent with the result by \cite{Du16a}. \cite{Du16a} found that $\dhb=2.01-0.29\log \mdot_{\rm F}$, and $\rfe=0.66+0.30\log \mdot_{\rm F}$. Fig. \ref{fig3} shows the distributions of \dhb and \rfe for three subsamples. The SEAMBHs subsample has smaller mean value of \dhb and has larger mean value of \rfe than other subsamples, which are consistent with their prevalently wide wing and strong \feii for SEAMBHs sources. Adopting $\mdot_{\rm F}$, these correlations are stronger than adopting $\mdot_{\rm \sigma}$ or $\mdot_{\rm c}$ for the entire sample or the sample excluding SDSS-RM AGNs. Adopting $\mdot_{\rm c}$, the relation between $\mdot_{\rm c}$ and \rfe is weakest. $\mdot_{\rm c}$ has a stronger correlation with \dhb than \rfe. It is the same as $\mdot_{\rm F}$. However, $\mdot_{\rm \sigma}$ has a weaker correlation with \dhb than \rfe (see values in top left corners in Fig. \ref{fig2}). For the entire sample, the relation of \mdot with \rfe or \dhb becomes weaker than for the sample excluding SDSS-RM AGNs. The SDSS-RM subsample has the larger scatter in all the correlations, with tighter correlations when the SDSS-RM subsample is excluded. Considering the observational uncertainties, this remains true. 

\subsection{Multivariate liner regression technique}
The multivariate regression analysis technique is used to find an extended \RL relation including other second parameter in the form: $ y = \alpha_{1}x_{1} + \alpha_{2}x_{2} + \beta_1$. We use the $\chi^2$ as the estimator to find the best values for these fitting parameters \citep{Me03}:
 \begin{equation}
\chi^2 = \Sigma_{i} \frac{(y_{i} - \alpha_{1}x_{1i}- \alpha_{2}x_{2,i}-\beta_1)^2}{\sigma_{\rm int}^{2}+\sigma_{y_{i}}^{2}+(\alpha_{1}\sigma_{x_{1i}})^2+(\alpha_{2}\sigma_{x_{2i}})^2},
 \label{eq5}
\end{equation}
where $y_{i}$ is the dependent variable. $x_{1i}$ and $x_{2i}$ are the independent variables. $\sigma_{y_{i}}$, $\sigma_{x_{1i}}$ and $\sigma_{x_{2i}}$ are the uncertainties of
$y_{i}$, $x_{1i}$ and $x_{2i}$, respectively. $\sigma_{\rm int}$ is the intrinsic scatter. Because Equation \ref{eq5} is non-linear in $\alpha_{1}$ and $\alpha_{2}$, it is impossible to minimize $\chi^2$ analytically. But for a given set of $\alpha_{1}$ and $\alpha_{2}$, we can solve the equation $\frac{\partial\chi^2}{\partial\beta_1}$ = 0. Therefore the optimal value $\beta_{1,\rm min}$
is
\begin{equation}
\label{eq6}
\beta_{\rm1, min}=\frac{\Sigma_{i}\frac{y_{i}-\alpha_{1}x_{1i}-\alpha_{2}x_{2i}}{\sigma_{\rm int}^2+\sigma_{y_{i}}^2+(\alpha_{1}\sigma_{x_{1i}})^2+(\alpha_{2}\sigma_{x_{2i}})^2}}{\Sigma_{i}
(\sigma_{\rm int}^2+\sigma_{y_{i}}^2+(\alpha_{1}\sigma_{x_{1i}})^2+(\alpha_{2}\sigma_{x_{2i}})^2)^{-1}},
\end{equation}
%For $\alpha_{1}$ and $\alpha_{2}$, we use a graphical solution in the form of 2-D $\chi^2$ contour plots.
The intrinsic scatter $\sigma_{\rm int}$ in Equation \ref{eq5} is  allowed to vary as an effective additional y error by repeating the fit until one obtains $\chi^{2}_{r}=1$ where $\chi^2_{r}=\chi^2/
N_{dof}$, $N_{dof}$ is the number of degree of freedom. The value of $\sigma_{\rm int}$ can derived by iteration \citep{Bam06, Bed06, Pa12}. In the first step set the $\sigma_{\rm int} = 0$ and compute the $\chi^2_{r}$. If $\chi^2_{r} \leq 1$ then $\sigma_{\rm int} = 0$. If $\chi^2_{r} \ > 1$ then set a initial value for the $\sigma_{\rm int,1}$ and do the iteration \citep{Bam06, Bed06, Pa12} :
\begin{equation}
\label{eq7}
\sigma^2_{\rm int,j+1}=\sigma^2_{\rm int,j} \chi_{r}^{2\alpha}, \alpha > 0.
\end{equation}
In this iteration, we adopt $\alpha = 2/3$. When $\chi_{r}^{2}  \ > 1$, $\sigma^2_{\rm int,j}$ is too small and thus should be increased for the next iteration. When  $\chi_{r}^{2}  \ < 1$, $\sigma^2_{\rm int,j}$ is too large and should be decreased for the next iteration.
When $\chi_{r}^{2}  = 1$ , $\sigma^2_{\rm int,j+1} = \sigma^2_{\rm int,j}$ and the value of $\sigma_{\rm int,j}$ is what we want.

The bootstrap method is used to estimate the error bars of $\alpha_{1}$, $\alpha_{2}$, $\beta$ and $\sigma_{\rm int}$. We resampled the data for 100 times and repeated the multiple linear regressions like above. Therefore we can derive 100 values for each parameters. For each parameter, we sort these values, adopt the 16th and 84th values as endpoints of the $1 \sigma$ confidence interval, and calculate its error.

\subsection{An extended \RL relation including the dimensionless accretion rate \mdot}

%\subsection{The canonical \RL relation}  %canonical

\begin{figure*}
\includegraphics[angle=-90,width=6.0in]{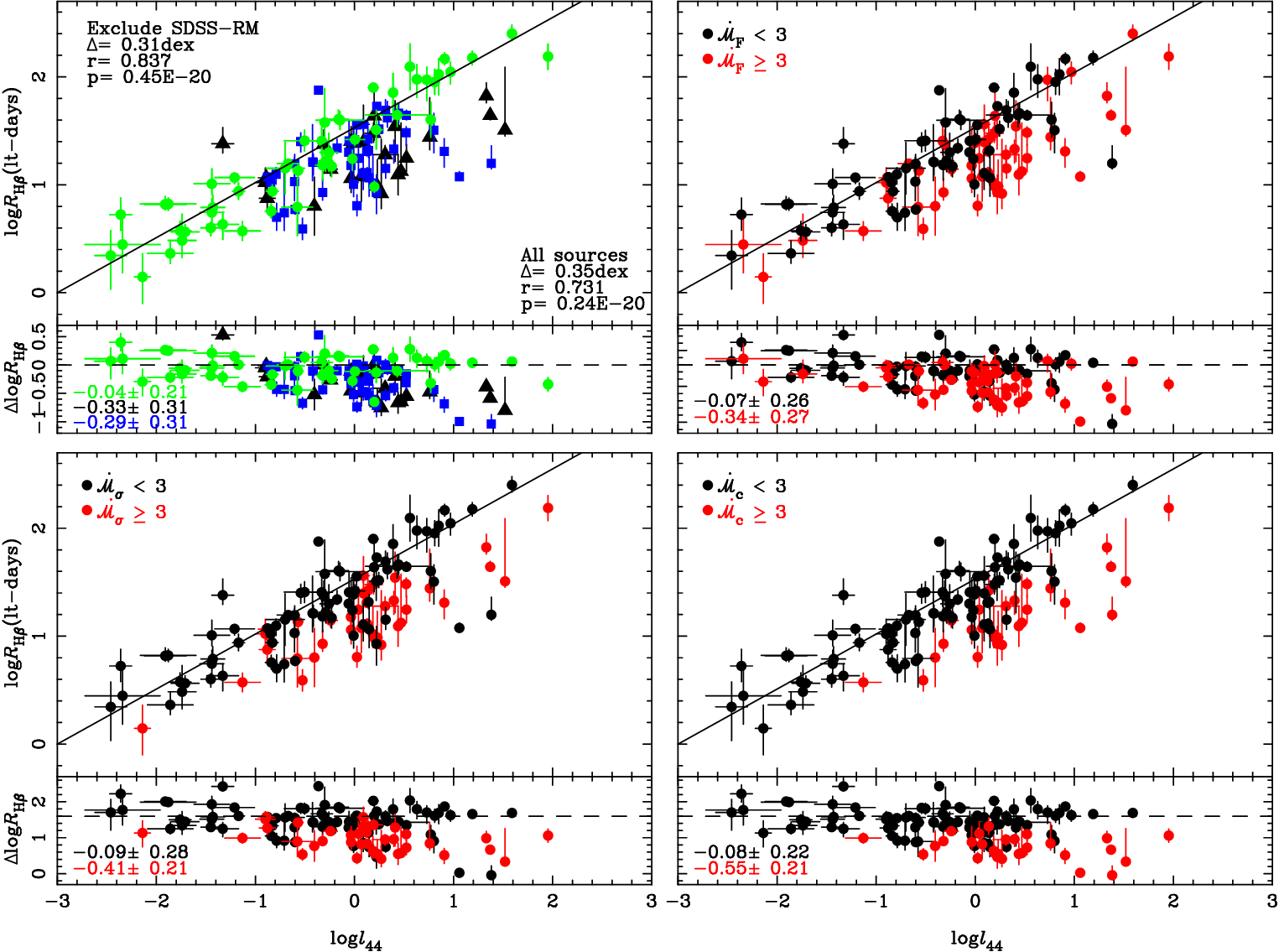}
\caption{ Left top: the $R_{\rm BLR}({\rm H\beta}) - l_{44}$ relation for BentzSample (green circles), SEAMBH (black triangles) and SDSS-RM (blue squares). The solid line corresponds to the $R_{\rm BLR}({\rm H\beta}) - l_{44}$ relation from Equation \ref{eq2} and $\alpha = 33.88$ light days and $\beta = 0.51$ \citep{Du18} for super-Eddington AGNs. The rms respect to the solid line, the Spearman correlation coefficient $r_s$ and probability of the null hypothesis $p_{\rm null}$ are shown in the upper left/right corner. The values of the mean and the standard deviation are shown in the left corner. Right top: the same \RL relation as the left panels for super-Eddington AGNs (red circles) and sub-Eddington AGNs (black circles) with the criterion of $\mdot_{\rm F} \ge 3$. Left bottom: the same as the top right panel but for $\mdot_{\rm \sigma}$. Right bottom: the same as the top right panel but for $\mdot_{\rm c}$.}
\label{fig4}
\end{figure*}

Because the SDSS-RM lags are computed from JAVELIN, which typically gives smaller uncertainties than the ICCF lags used in the other two subsamples \citep[e.g.][]{Yu19},
we simply use the canonical \RL relation derived by sub-Eddington AGNs for our sample excluding SDSS-RM AGNs with the fibre spectra. Fig. \ref{fig4} shows the \RL relation for our sample. The solid line is defined by the Equation \ref{eq2}, where $\alpha = 33.88$ light days and $\beta = 0.51$ for sub-Eddington AGNs ($\mdot_{\rm F} < 3$) from our sample excluding SDSS-RM AGNs, which is almost the same as \cite{Du18}. We adopted it as the canonical \RL relation. For our sample excluding SDSS-RM AGNs, the Spearman correlation coefficient $r_s = 0.837$, the probability of the null hypothesis $p_{\rm null} = 0.45 \times 10^{-20}$, and the offset rms (with respect to the canonical \RL relation) is 0.31 dex. For our entire sample, $r_s=0.731$, $p_{\rm null}=0.24\times 10^{-20}$, and the offset rms is 0.35 dex. For subsamples of SEAMBH AGNs and SDSS-RM AGNs, there is a downward offset with respect to the solid line. Using $\mdot_{\rm F}$, the mean value of the offset from the canonical \RL relation is respectively  -0.36 dex, -0.16 dex, -0.48 dex, and -0.34 dex for super-Eddington accretion SMBHs in three subsamples (SEAMBHs, BentzSample, SDSS-RM), and entire sample. Using $\mdot_{\rm \sigma}$, the values are -0.42 dex, -0.37 dex, -0.42 dex, -0.41 dex, repevtively. Using $\mdot_{\rm c}$, the values are -0.55 dex, -0.46 dex, -0.59 dex, -0.55 dex, respectively. It is found that super-Eddington AGNs has a negative mean offset with respect to the canonical \RL relation for sub-Eddington AGNs, no matter what kind of \mdot.
Particularly for the BentzSample subsample, the offset of 14/51 super-Eddington accretion SMBHs is smallest ($-0.16\pm 0.28$ dex, statistically marginal) for the simple FWHM-based $\mdot_{\rm F}$, and there is only three super-Eddington AGNs with offset of -0.46 dex for the corrected-FWHM $\mdot_{\rm c}$.

%Including additional sub-Eddington AGNs from SDSS-RM subsample, we also do the linear regression to obtain the canonical \RL relation instead of the relation by \cite{Du18}, and find a consistent result with the above one. 

\begin{figure*}
\includegraphics[angle=-90,width=6.3in]{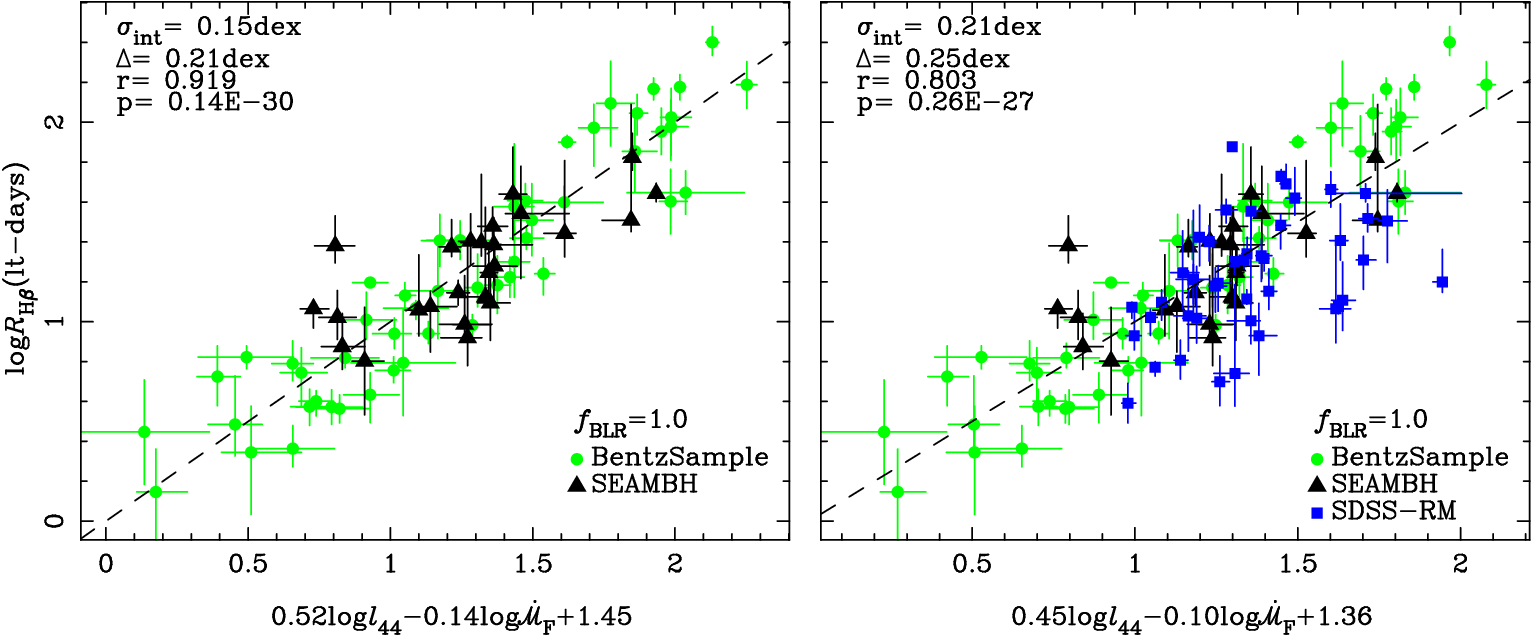}
\includegraphics[angle=-90,width=6.3in]{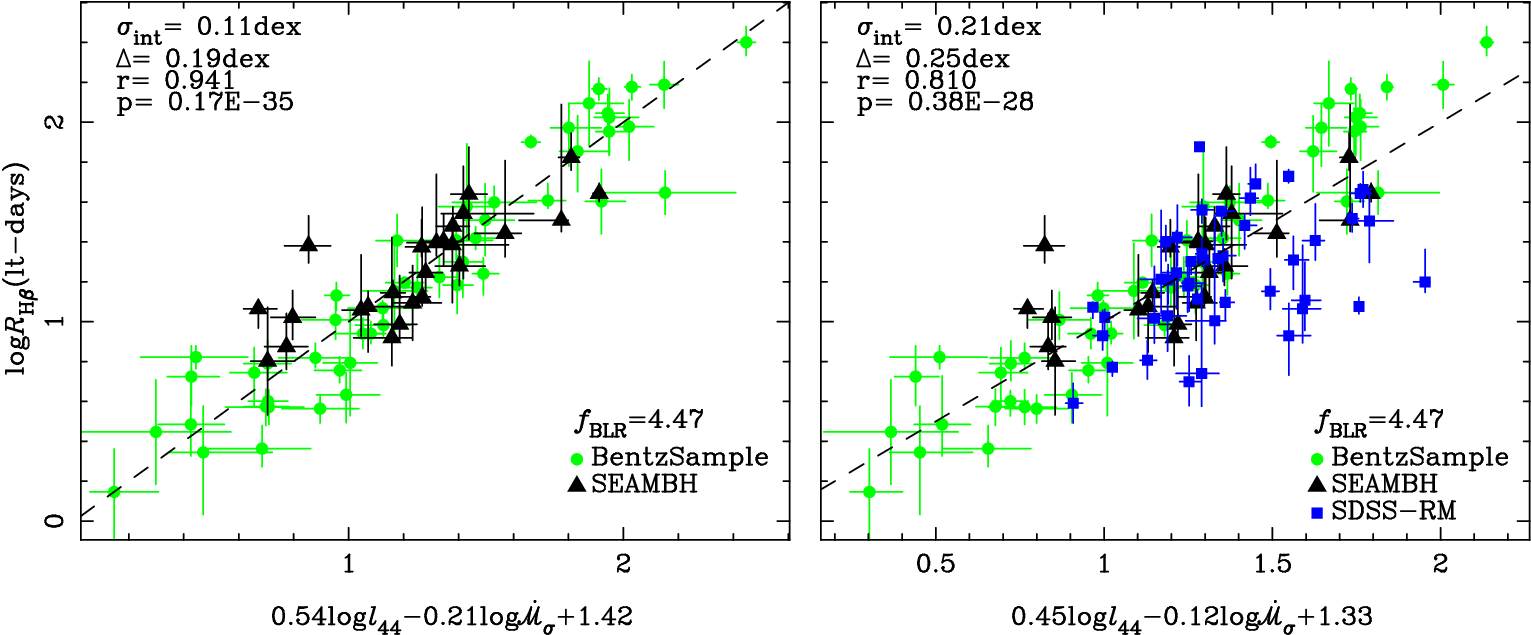}
\includegraphics[angle=-90,width=6.3in]{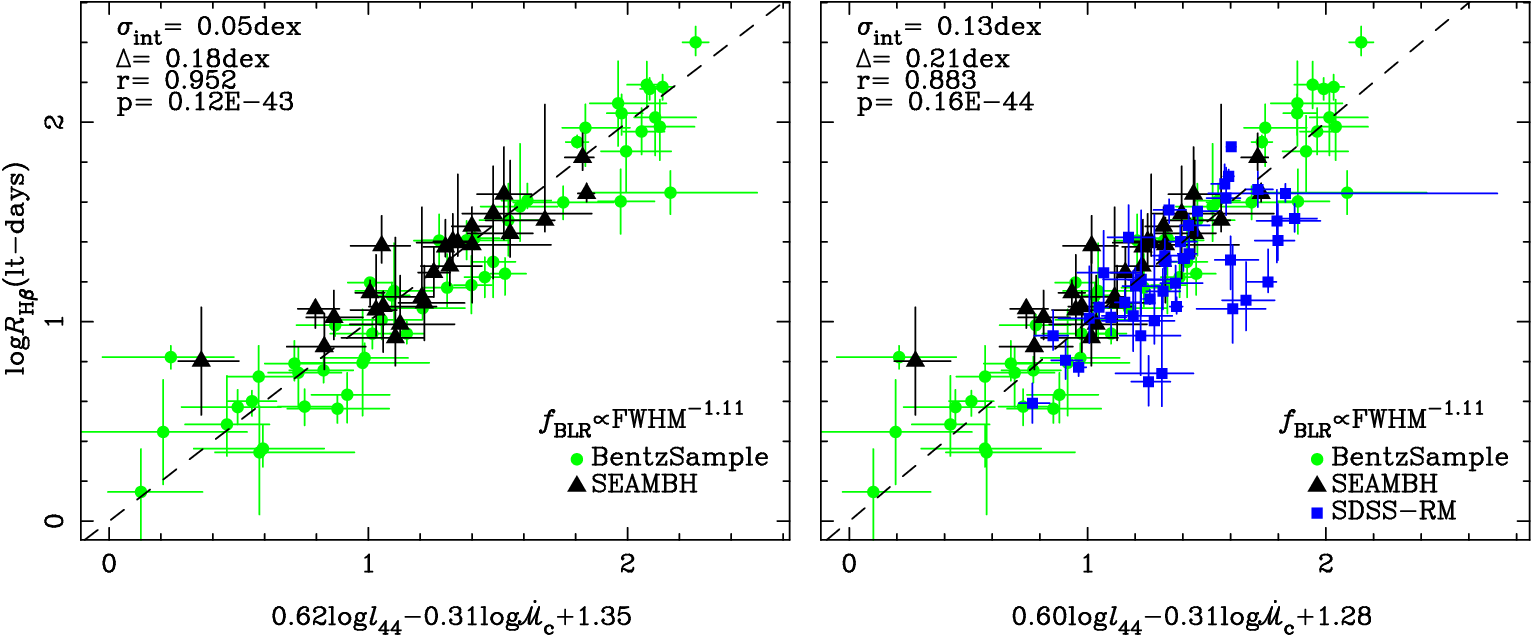}\vfill
\caption{Dependence of the \rhb on the $l_{44}$ and the \mdot. The left panels demonstrate the dependence for all the RM AGNs excluding the SDSS-RM AGNs. The right panels demonstrate the dependence for all the RM AGNs. The symbols are the same to the Fig. \ref{fig2}. The dash line is 1:1. In the top two panels, the dimensionless accretion rate is derived from FWHM and the virial factor $f_{\rm BLR}$=1. In the middle two panels, the dimensionless accretion rate is derived from \shb and $f_{\rm BLR}$=4.47. In the bottom two panels, the dimensionless accretion rate is derived from FWHM and the virial factor $f_{\rm BLR} \propto \rm FWHM^{-1.11}$. The intrinsic scatter, the offset rms respect to the solid line, the Spearman correlation coefficient $r_s$ and probability of the null hypothesis $p_{\rm null}$ are shown in the left corner in each panel. }
\label{fig5}
\end{figure*}

%In Fig.\ref{fig3}, we noticed that the SEAMBH sources indeed deviate from the canonical \RL relation \citep{Be13, Du16b}.

Considering the deviation $\Delta R_{\rm BLR}(\rm \hb)$ from the canonical \RL relation defined by sub-Eddington AGNs, it was suggested that objects with \mdot $<$ 3 would be associated with $\Delta R_{\rm BLR}(\rm \hb) \sim$ 0, while objects with \mdot $\ge$ 3 the deviation would decrease as a function with \mdot \citep{Du16b, Du18}. Since some super-Eddington AGNs deviate from the canonical \RL relation, it is necessary to give a new extended \RL relation. Considering \mdot as the key parameter \citep{Du18,Ma2019}, we use the multivariate liner regression technique as described in Section 3.2 to find an extended \RL relation, i.e., the correlation between $R_{\rm BLR}(\rm \hb)$, \lv and \mdot. For three kinds of \mdot and excluding SDSS-RM or not, we perform the bivariate liner regressions respectively and present the results in Table \ref{tab2}. Including the dimensionless accretion rate \mdot, the relations in Fig. \ref{fig5} become stronger (see Table \ref{tab2}). 

In Fig. \ref{fig5}, we present the dependence of the $R_{\rm BLR}(\rm \hb)$ on the \lv and \mdot. Three top panels are for our sample excluding the SDSS-RM sources and three bottom panels are for all sources. In two left panels, \mdot is derived from the \hb FWHM and $f_{\rm BLR}$ = 1. In two middle panels, \mdot is derived from \shb and $f_{\rm BLR}$=4.47. In two right panels, \mdot is derived from the \hb FWHM and the FWHM-based factor $f_{\rm BLR} \propto \rm FWHM^{-1.11}$. When the SDSS-RM sources are included, the significance of the correlation decreases (shown in right panels in \ref{fig5}). Including \mdot in the \RL relation, the correlations become stronger than the canonical \RL relation and smaller values of the offset rms (see Fig. \ref{fig4}), especially for $\mdot_{\rm c}$ derived from our virial factor calibration \citep{Yu19}. Adopting $\mdot_{\rm c}$, for excluding the SDSS-RM AGNs, the intrinsic scatter is $\sigma_{\rm int}$ = 0.05 dex,  $R_{\rm BLR}$ offset rms (with respect to this relation) is 0.18 dex, $r_s=0.952$ and $p_{\rm null} = 0.12\times 10^{-43}$. For all RM AGNs, $\sigma_{\rm int}$ is 0.13 dex, the $R_{\rm BLR}$ offset rms is 0.21 dex, $r_s = 0.883$ and $p_{\rm null}=0.16\times 10^{-44}$. Adopting $\mdot_{\rm c}$, an order of magnitude in $\mdot_{\rm c}$ would lead to a change of 0.31 dex in $\log R_{\rm BLR}$ for the bivariate relation shown in the bottom left panel shown in Fig. \ref{fig5}. Adopting $\mdot_{\rm F}$ or $\mdot_{\rm \sigma}$, an order of magnitude in \mdot would lead to a change of 0.14 dex or 0.21 dex in $\log R_{\rm BLR}$ for the bivariate relation shown in the top left panel or middle left panel shown in Fig. \ref{fig5}. There are large range of \mdot for AGNs from hot accretion to standard accretion to super-Eddington accretion \citep{YN2014}. Therefore, the role of \mdot in the \RL relation would be important (see additional discussion in section 4).

\subsection{An extended \RL relation including the observational parameter \rfe or \dhb}

\begin{figure*}
\includegraphics[angle=-90,width=6.29in]{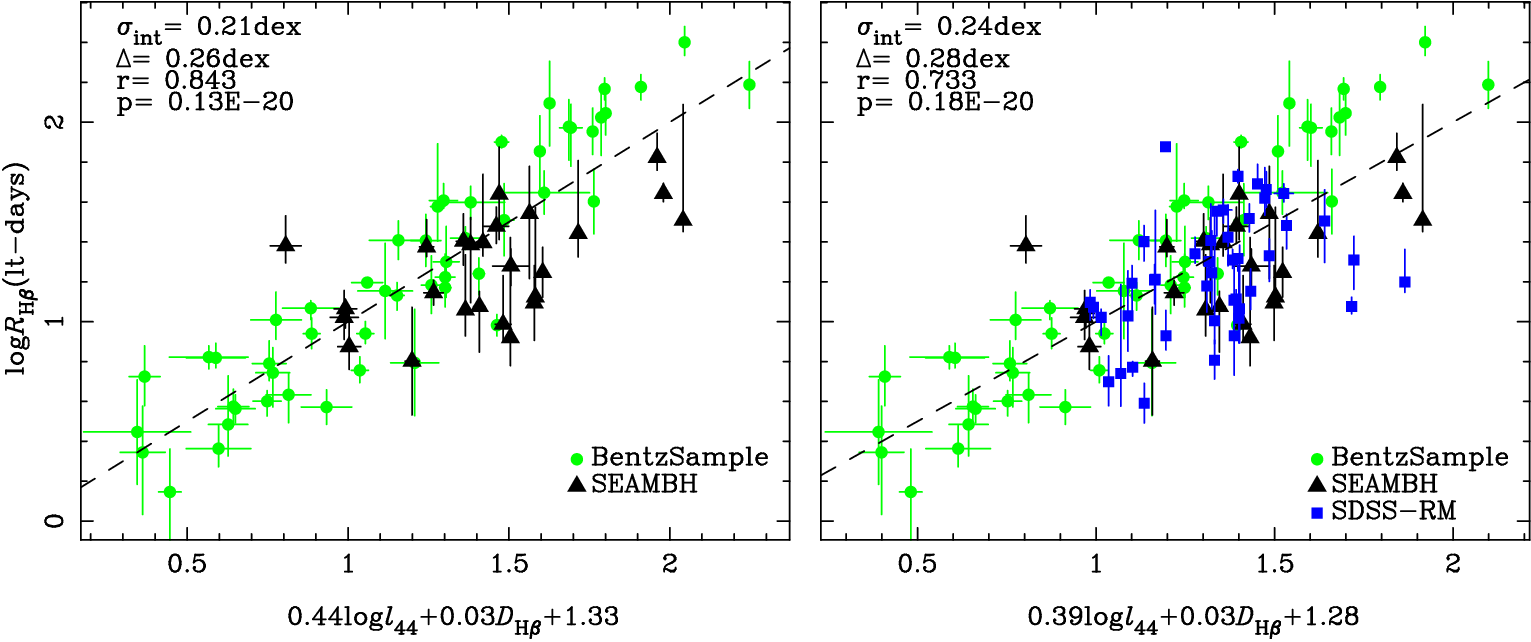}
\includegraphics[angle=-90,width=6.3in]{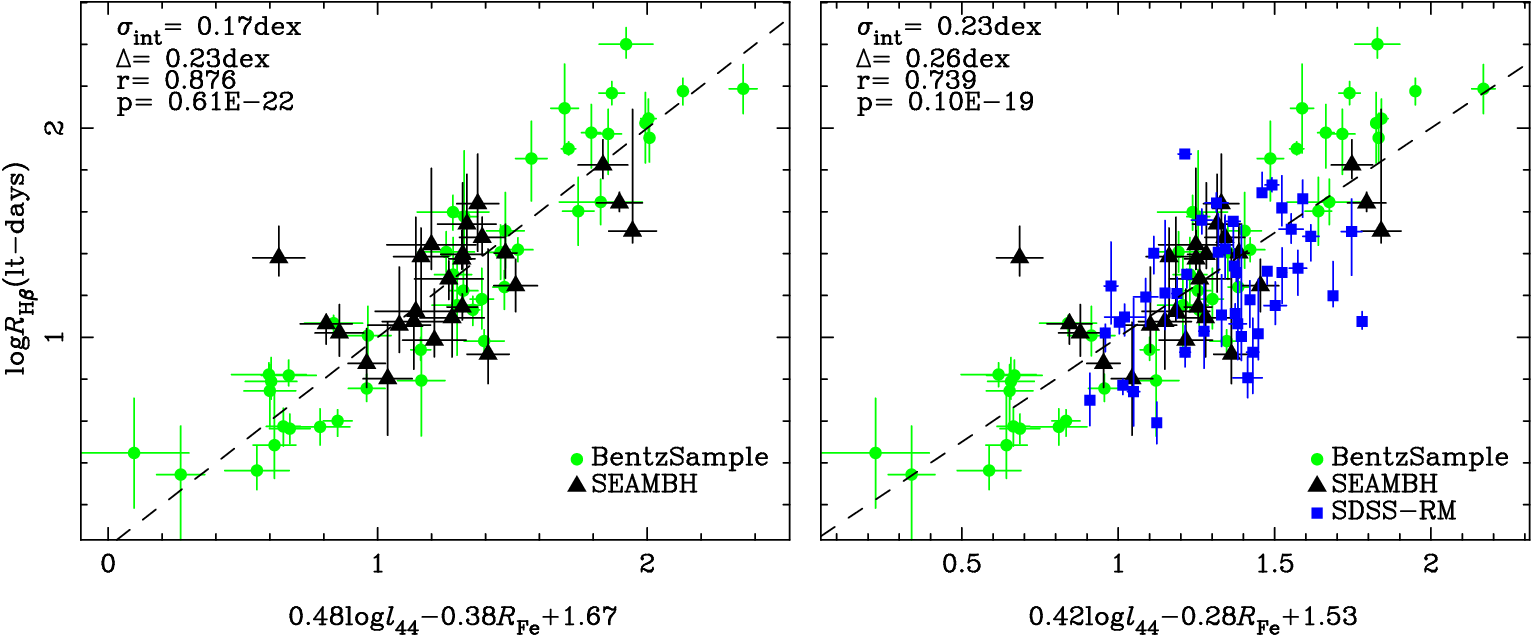}
\caption{Dependence of the \rhb on the $l_{44}$ and the \dhb (top). Dependence of the \rhb on the $l_{44}$ and the \rfe (bottom). Two left panels demonstrate the dependence for all the RM AGNs excluding the SDSS-RM AGNs and two right panels demonstrate the dependence for all the RM AGNs. The symbols are the same to the Fig. \ref{fig2}. The dash line is 1:1. The intrinsic scatter, the offset rms respect to the solid line, the Spearman correlation coefficient $r_s$ and probability of the null hypothesis $p_{\rm null}$ are shown in the left corner in each panel.}
\label{fig6}
\end{figure*}

The dimensionless accretion rate \mdot is computed directly from the continuum $L_{5100}$  and also (for most sources) from $R_{\rm BLR}$ (see equation \ref{eq3}). This means that any comparison of \mdot with the other quantities suffers from self-correlation. \rfe and the \hb shape \dhb are measures of \mdot that are independent of $R_{\rm BLR}$ and $L_{5100}$. These two observational parameters of \rfe and \dhb can be measured from the single-epoch spectrum.
In section 3.1, it is found that \rfe or \dhb has a relation with \mdot (see Fig. \ref{fig2}). Therefore, we use the multivariate regression analysis technique to investigate the extended \RL relation including observational parameters, i.e., \rfe and \dhb.
In Fig. \ref{fig6}, new empirical \RL relations are presented including \rfe or \dhb. Excluding SDSS-RM AGNs, the extended \RL relations are
\begin{equation}
\label{eq8}
\begin{split}
\log \rhb = & 
 (0.48\pm 0.03) \log l_{44}  -(0.38\pm 0.04) \rfe \\ & +(1.67\pm 0.09)~~\rm ltd, \\
\log \rhb = &
 (0.44\pm 0.03) \log l_{44} - (0.03\pm 0.04) \dhb \\ & +(1.33\pm 0.08)~~\rm ltd.
\end{split}
\end{equation}

For the subsample of excluding SDSS-RM AGNS, in an extended \RL relation including \rfe, $r_s=0.876$, $p_{\rm null}=0.61\times 10^{-22}$. Including \dhb, $r_s=0.843$, $p_{\rm null}=0.13\times 10^{-20}$. Including \rfe or \dhb in the \RL relation, $r_s$ increase with respect to that for the \RL relation ($r_s=0.837$, $p_{\rm null}=0.45\times 10^{-20}$) shown in Fig. \ref{fig4}. The offset rms decreases from 0.31 dex to 0.26 dex (including \dhb) or 0.23 dex (including \rfe).
For the entire sample, in this extended relation including \rfe or \dhb, $r_s=0.733$ (for \dhb) or $r_s=0.739$ (for \rfe), which is slightly larger than $r_s=0.731$ for the \RL relation for the entire sample (see Fig. \ref{fig4}). The offset rms decreases from 0.35 dex to 0.28 dex (including \dhb) or 0.26 dex (including \rfe). 

Although the extended \RL relations become weaker for the entire sample than for the subsample excluding SDSS-RM AGNs, the extended \RL relations including \rfe and \dhb become stronger than for the \RL relation shown in Fig. \ref{fig4}. Including \rfe, the improvement of \RL relation is more significant than for including \dhb (the y-axis offset rms is 0.23 dex versus 0.26 dex). 
%A change of 1 in \dhb would lead to a change of 0.06 dex in $\log R_{\rm BLR}$ for the bivariate relation shown in the top left panel in Fig. \ref{fig6}. 
A change of 1 in \rfe would lead to a change of 0.38 dex in $\log R_{\rm BLR}$ for the bivariate relation shown in the bottom left panel in Fig. \ref{fig6}.  
Considering a slope of $0.03\pm 0.04$ for the \hb shape \dhb, it should be noted that the best-fit relation with \dhb shows only a marginal dependence on this quantity (i.e., a slope that \~ 1 $\sigma$ different from zero). \rfe is a better "fix" for the \RL offset than the \hb shape \dhb. Including SDSS-RM AGNs, a change of 1 in \rfe would lead to a smaller change of $\log R_{\rm BLR}$, from 0.38 dex to 0.28 dex. \rfe has more contribution in \RL relation than for \dhb. It is consistent well with the result by \cite{Du19}. Considering the correlations of $\mdot_{\rm F}$ with \rfe or \dhb found by \cite{Du16a}, from Eq. \ref{eq8} we can derive that $R_{\rm BLR} \propto l_{44}^{0.49}~ \mdot_{\rm F}^{0.114}$, or $R_{\rm BLR} \propto l_{44}^{0.45} ~ \mdot_{\rm F}^{0.023}$. They have some slight difference comparing with the equations shown in the top left panel in Fig. \ref{fig5}. It is due to large scatter when using \rfe or \dhb to substitute \mdot. 

\section{Discussion}
\subsection{The \RL relation: scatter and origin }

Considering the photonionization model of the BLRs in AGNs \citep{FN83},  the ionization parameter $U=\frac{Q(\rm H)}{n_e 4\pi R^2c}$, where R is the BLRs size  for the \hb broad line, c is the speed of light, $n_e$ is the electron number density, and $Q(\rm H)=\int^{\infty}_{13.6eV} \frac{L_{\nu}}{h\nu} ~d \nu$ is the flux of hydrogen ionizing photons emitted by the central source. The ionization level of the BLR clouds can be estimated using either the ionizing flux $L_{\rm ion}$ at 912\AA, or $Q$ \citep[e.g.][]{Wang2014a, C2019}. Adopting some assumptions, such as the point central ionization source, the same $U$, the same $n_e$, the same SED, and the same BLRs geometry, we can derive $R_{\rm BLR} \propto Q(\rm H)^{0.5} \propto L_{5100}^{0.5}$, i.e., the \RL relation. The  UV/optical luminosity ratio depends on detail SMBH accretion process, which can be used to explain the shorter lags than that expected from the canonical \RL relation \citep{Wang2014a, C2019}. It is also possible for different values of  $U$ or $n_e$ for different AGNs. 
%In fact, the relation between \lv and $Q(\rm H)$ or $L_{\rm ion}$ is non-linear and depends on SED. 

The change in SED with the luminosity is suggested that brighter AGNs have flatter optical-UV continua, ranging from an average slope ($\alpha$ in $f_{\nu}\propto \nu^{\alpha}$) of about -1 for Seyfert galaxies to about -0.3 for quasars (although host-correction is a problem) \citep[e.g.][]{Ka05}. Quasars have a larger UV/optical luminosity ratio than Seyfert galaxies. Using UV luminosity to substitute the optical luminosity, the slope of the BLR size versus luminosity relation becomes shallower \citep{Ka05}. \cite{Ho2008} also gave the different slope of the optical-UV continuum binned by the Eddington ratio, and found large UV/optical luminosity ratio for nearby AGNs with large Eddington ratios (see their Fig. 7). For our result of the \mdot role in the \RL relation, larger \mdot leads to smaller $R_{\rm BLR}$ at fixed \lv\ (see Fig. \ref{fig5}). AGNs with large luminosity or large \mdot always have large UV/optical luminosity ratio, which leads to a relatively large BLR size. Therefore, it is difficult to explain the smaller $R_{\rm BLR}$ at fixed \lv, especially for bright SEAMBHs.

For different SMBH spins, \mbh, and \mdot, the theoretical UV/optical luminosity ratio were calculated from the general relativistic version of the \cite{SS73} disk \citep{Wang2014a, C2019}. For AGNs with larger Eddington ratio, smaller \mbh, large spin, the relation between $L_{\rm ion}$ and \lv is almost linear. For other conditions, the temperature of the accretion disk drops to make the SED peak moving into UV band, leading to a a non-linear relation between $L_{\rm ion}$ and \lv. Large black hole mass, low Eddington ratio and low spin lead to significant differences in the slope of the optical-UV continuum \citep{Ho2008, Trump2011,Lusso2012}.
%curvature of the spectrum in the UV band. 
The retrograde spin leads to a cold accretion disk and makes large scatter in the \RL relation. Therefore, for the smaller $R_{\rm BLR}$ at fixed \lv, the reason is small Eddington ratio or small SMBH spin, which lead to a cold disk, and a deficit of ionizing photons in the BLR. For SEAMBHs with largest \mdot, the self-shadowing effects from a slim disk would lead to a deficit of ionizing photons illuminating BLR clouds \citep{Wang2014a,Wang2014b}. Therefore, the SED changes, connected to changes in \mdot, are the most likely culprit for the breadth in the \RL relation in the first-order consideration.

It is also noted that there are some scatter in our extended \RL relation including \mdot. In Fig. \ref{fig5} and  Fig. \ref{fig6}, it is found that the correlations become weaker when including SDSS-RM quasars. There exists the excess \RL scatter of the SDSS-RM quasars. The SDSS-RM subsample is less well explained by the extended \RL relations than are the subsamples of BentzSample and SEAMBHs. One reason is that the SDSS-RM lags are computed from JAVELIN, which typically gives smaller uncertainties than the ICCF lags used in the other two samples. The other is possibly due to the luminosity-threshold \citep{Sh15} or the retrograde spin for SDSS-RM subsample \citep{Du18}.

Our result of the dependence of the \RL relation on \mdot supports that $R_{\rm BLR}$ has a relation with the accretion process. Using the multivariate regression analysis technique, for the  entire sample, we also find the relation of $R_{\rm BLR}$ with \mbh and \mdot, i.e., $R_{\rm BLR} \propto \mbh^{0.61} \mdot_{\rm F}^{0.20}$ ($r_s=0.774$), $R_{\rm BLR} \propto \mbh^{0.61} \mdot_{\sigma}^{0.18}$ ($r_s=0.787$), $R_{\rm BLR} \propto \mbh^{0.85} \mdot_{\rm c}^{0.09}$ ($r_s=0.876$). For a simple model of gravitational instability of the standard accretion disk leading the formation of the BLRs clouds \citep{bian2002}, $R_{\rm  BLR} \propto \lv^{0.5} \dot{M}^{-37/45} \propto \mbh^{-7/45} \mdot^{-22/45}$. Therefore, our measured accretion rate dependence is not consistent with the simple model of the gravitational instability of accretion disk as the origin of the BLRs. It was found that the size of torus dust is about 4 times of the BLRs size from the optical/near-infrared RM \citep[e.g.][]{Kokubo2019}. The BLR may instead form from the inner edge of the torus\citep[e.g.][]{Wang2017}, or from some other means in which BLR size is positively correlated with accretion rate and the SMBH mass.

\subsection{The extended empirical \RL relation: \rfe or \dhb}
It was found that EV1 is driven by the accretion rate \citep[e.g.][]{Boroson2002}. From the optical spectra of AGNs, we use two EV1-related parameters of \rfe and \dhb to substitute \mdot. In Fig. \ref{fig2}, both \rfe and \dhb seem to effectively correlate with \mdot. From Fig. \ref{fig3}, \rfe has a wide distribution than \dhb, which is around 2. For the distribution of \rfe, the ratios of the mean value to the rms is 3.03, 1.92, 1.77, respectively for three subsamples of SEAMBHs, SDSS-RM, BentzSample. For the distribution of \dhb, the ratios of the mean value to the rms is 10.93, 3.36, 3.17, respectively for these three subsamples. For the subsample of SEAMBHs, the \rfe has a wide distribution and \dhb has a narrow distribution. The wide distribution of \rfe than \dhb possibly leads to that the \hb shape \dhb less correlates with \RL offset than \rfe (see Fig. \ref{fig6}). \rfe is a better "fix" for the \RL offset than \dhb. This empirical \RL relation including \rfe would be used to derive the BLR size for the \hb broad line instead of the canonical \RL relation, and then weigh \mbh from the single-epoch spectrum.

\section{Conclusions}
For a compiled sample of 120 RM AGNs, the dependence of the \RL relation on \mdot is investigated. Using observational parameters of \rfe and \dhb, extended empirical $R_{\rm BLR}-L_{\rm 5100}$ relations are presented, which can be used to calculate the BLRs sizes from a single-epoch spectrum. The main conclusions are summarized as follows:

\begin{itemize}
\item Using our recently calibrated virial factor $f$, and the velocity tracer from the \hb FWHM or the line dispersion \shb from the mean spectrum, three kinds of SMBH masses \mbh and the dimensionless accretion rates \mdot are calculated for a large compiled sample of 120 RM AGNs. The classification of super-Eddington AGNs depends on the calculation method of \mbh and thus \mdot. It is found that \rfe or \dhb has a relation with \mdot. 

\item  Including the effect of \mdot, the bivariate correlation of $R_{\rm BLR}$ with \lv and \mdot has a smaller scatter than that for the canonical \RL relation. The correlation coefficient $r_s$ is larger and $p_{\rm null}$ is smaller than the canonical \RL relation. For \mbh and \mdot derived from \hb FWHM and \hb FWHM-based f, the bivariate correlation of $R_{\rm BLR}$ with $L_{5100}$ and $\mdot_{\rm F}$ has a smallest scatter for three cases of \mdot. 

\item Substituting observational parameter of \rfe or \dhb for \mdot, extended empirical \RL relations are found, which would be used to derive $R_{\rm BLR}$ instead of by the canonical \RL relation, and then \mbh from the single-epoch spectrum. Including the optical \feii relative ratio \rfe, the improvement of \RL relation is more significant than for including the line profile parameter \dhb. \rfe is a better "fix" for the \RL offset than the \hb shape \dhb. The best relation is $\log R_{\rm BLR}=(0.42\pm 0.03)\log l_{44}-(0.28\pm 0.04)\rfe+(1.53\pm 0.08)~\rm ltd$ for the entire sample, with an intrinsic scatter of 0.23 dex.

\item Although our measured accretion rate dependence is not consistent with the simple model of the accretion disk instability leading the BLRs formation, our results show that the accretion rate has an important role in the \RL relation. The BLR may instead form from the inner edge of the torus, or from some other means in which BLR size is positively correlated with accretion rate and the SMBH mass.

\end{itemize}

\section*{Acknowledgements}
We are also very grateful to the anonymous referee for her/his instructive comments which significantly improved the content of the paper.
This work is supported by the National Key Research and Development Program of China (No. 2017YFA0402703). This work has been supported by the National Science Foundations of China (Nos. 11973029 and 11873032).

\newpage
\begin{table*}
\caption{The properties of 120 RM AGNs.}
\label{tab1}
\begin{lrbox}{\tablebox}
\begin{tabular}{llllllllllllll}
\hline
Name&$\tau$&$\log \lv$&FWHM&\shb&\dhb&\rfe&$\log M_{\rm BH,F}$&$\log M_{\rm BH, \sigma}$&$\log M_{\rm BH, c}$&$\log \mdot_{F}$&$\log \mdot_{\sigma}$&$\log \mdot_{c}$&Ref\\
         &(days)&$\log \ergs$&(\kms)&(\kms)&&&($\log \msun$)&($\log \msun$)&($\log \msun$)&&&\\
 (1)&(2)&(3)&(4)&(5)&(6)&(7)&(8)&(9)&(10)&(11)&(12)&(13)&(14)\\
\hline
\multicolumn{14}{c}{SEAMBH}\\
\hline
Mrk335&                  $  8.7_{-  1.9}^{+  1.6}$&$ 43.69\pm  0.06$&$2096 \pm170 $&$1470 \pm50  $&$  1.43\pm 0.09$&        $\cdots$&$ 6.87_{- 0.13}^{+ 0.10}$&$ 7.21_{- 0.11}^{+ 0.08}$&$ 7.33_{- 0.13}^{+ 0.10}$&$ 0.53_{- 0.13}^{+ 0.13}$&$ 0.53_{- 0.13}^{+ 0.13}$&$ 0.53_{- 0.13}^{+ 0.13}$&$1, 2, 3^{\ast}$       \\
Mrk335&                  $ 16.8_{-  4.2}^{+  4.8}$&$ 43.76\pm  0.06$&$1792 \pm3   $&$1380 \pm6   $&$  1.30\pm 0.00$&        $\cdots$&$ 7.02_{- 0.12}^{+ 0.11}$&$ 7.45_{- 0.12}^{+ 0.11}$&$ 7.55_{- 0.12}^{+ 0.11}$&$ 0.63_{- 0.13}^{+ 0.13}$&$ 0.63_{- 0.13}^{+ 0.13}$&$ 0.63_{- 0.13}^{+ 0.13}$&$2, 3, 4^{\ast}$       \\
\hline
\end{tabular}
\end{lrbox}
\scalebox{0.7}{\usebox{\tablebox}}
\\ 
Note. For a object with multiple measurements, the dimensionless accretion rate \mdot is calculate by the weighted average of the black masses. Names in boldface are the weighted averages of all the measurements.
$\ast$ means that the MCMC black hole mass is used to calculate the dimensionless accretion rate.
Reference: 1: \cite{Du15}, 2: \cite{Du16a}, 3:\cite{Co06}, 4:\cite{Be13}, 5:\cite{Gr12}, 6:\cite{Be09a},  7: \cite{Du18}, 8: \cite{Du16b},  9: \cite{Sh15},  10: \cite{Gr17},  11: \cite{Shen19},   12:  \cite{Be09b},  13: \cite{Denny2010},   14: \cite{Pe00},   15: \cite{Be06}, 16: \cite{Fau2017} 17: \cite{Zhang19}  18: \cite{Denny2006}   19: \cite{Be14},   20: \cite{Lu16},  21: \cite{Pei17},  22: \cite{Ba13} ,  23: \cite{Pei14},  24: \cite{Pe14}, 25:\cite{HK14},   26: \cite{Ba15},   27: \cite{Williams2018}, 28: \cite{Be16a},  29: \cite{Be16b}
\end{table*}

\begin{table*}
\caption{The results of of the multivariate liner regression. The $\alpha_{1}$, $\alpha_{2}$ and $\beta_1$ is defined by: $y = \alpha_{1}x_{1} + \alpha_{2}x_{2} + \beta_1$. The $\sigma_{\rm int}$ is the intrinsic scatter of this relation. The uncertainties of each value is derived from bootstrap simulation as described in Section 3.2. The last two column are the Spearman correlation coefficient $r_s$ and probability of the null hypothesis $p_{\rm null}$, we highlighted the correlations  with the significant value ($|r_s| \ >$ 0.85) in boldface. }
\begin{tabular}{lllrrcll}
\hline
                                                                  &                 & \multicolumn{1}{c}{$\alpha_{1}$} &  \multicolumn{1}{c}{$\alpha_{2}$} &  \multicolumn{1}{c}{$\beta_1$} & $\sigma_{\rm int}$&$r_s$&$p_{\rm null}$\\
\hline
\multirow{2}{*}{$\log \rhb=\alpha_{1}\log l_{44}+\alpha_{2}\log \mdot_{F}+\beta$}& All Sources&$   0.45\pm   0.03$&$  -0.10\pm   0.02$&$   1.36\pm   0.02$&$   0.21\pm   0.02$&  0.803& 0.26E-27\\
&Exclude SDSS-RM&$   0.52\pm   0.03$&$  -0.14\pm   0.02$&$   1.45\pm   0.02$&$   0.15\pm   0.02$&  \bfseries 0.919& 0.14E-30\\
\multirow{2}{*}{$\log \rhb=\alpha_{1}\log l_{44}+\alpha_{2}\log \mdot_{\sigma}+\beta$}& All Sources&$   0.45\pm   0.03$&$  -0.12\pm   0.02$&$   1.33\pm   0.02$&$   0.21\pm   0.02$&  0.810& 0.38E-28\\
&Exclude SDSS-RM&$   0.54\pm   0.03$&$  -0.21\pm   0.02$&$   1.42\pm   0.02$&$   0.11\pm   0.02$&  \bfseries 0.941& 0.17E-35\\
\multirow{2}{*}{$\log \rhb=\alpha_{1}\log l_{44}+\alpha_{2}\log \mdot_{c}+\beta$}& All Sources&$   0.60\pm   0.03$&$  -0.31\pm   0.03$&$   1.28\pm   0.02$&$   0.13\pm   0.02$& \bfseries 0.883&0.16E-44\\
& Exclude SDSS-RM&$   0.62\pm   0.02$&$  -0.31\pm   0.02$&$   1.35\pm   0.02$&$   0.05\pm   0.04$& \bfseries 0.952& 0.12E-43\\
\multirow{2}{*}{$\log \rhb=\alpha_{1}\log l_{44}+\alpha_{2}\dhb+\beta$}& All Sources&$   0.39\pm   0.02$&$   0.03\pm   0.04$&$   1.28\pm   0.09$&$   0.24\pm   0.02$&  0.733& 0.18E-20\\
& Exclude SDSS-RM&$   0.44\pm   0.03$&$   0.03\pm   0.04$&$   1.33\pm   0.08$&$   0.21\pm   0.02$&  0.843& 0.13E-20\\
\multirow{2}{*}{$\log \rhb=\alpha_{1}\log l_{44}+\alpha_{2}\rfe+\beta$}& All Sources&$   0.42\pm   0.03$&$  -0.28\pm   0.04$&$   1.53\pm   0.08$&$   0.23\pm   0.02$&  0.739& 0.10E-19\\
& Exclude SDSS-RM&$   0.48\pm   0.03$&$  -0.38\pm   0.04$&$   1.67\pm   0.09$&$   0.17\pm   0.02$& \bfseries 0.876& 0.61E-22\\
  \hline
\end{tabular}
\label{tab2}
\end{table*}

\end{document}